\documentclass{article}

\usepackage[utf8]{inputenc}
\usepackage[english]{babel}

\usepackage[backend=biber, style=phys, sorting=none, doi=false, biblabel=brackets]{biblatex}
\usepackage{csquotes}

\AtEveryBibitem{%
  \clearfield{Extra}%
}

\AtEveryBibitem{%
  \clearfield{note}%
}

\AtEveryBibitem{%
  \clearfield{series}%
}

\AtEveryBibitem{%
  \clearlist{language}%
}

\usepackage{graphicx}
\usepackage{float}
\usepackage{subfigure}
\usepackage{placeins}

\usepackage{indentfirst}

\DeclareBibliographyCategory{cited}
\AtEveryCitekey{\addtocategory{cited}{\thefield{entrykey}}}

\defbibheading{ref}{\newpage\centering References}
\defbibheading{supp}{\newpage\centering Supplementary References}

\usepackage{verbatim}
\usepackage{amsmath}

\usepackage{setspace}

\usepackage{lineno}

\usepackage[figurename=FIG., labelsep=period]{caption}
\usepackage[font=small,labelfont=bf]{caption}
\usepackage{authblk}

\usepackage[T1]{fontenc}
\usepackage{lmodern}
\usepackage{textalpha}
\DeclareUnicodeCharacter{2212}{-}
\DeclareUnicodeCharacter{2009}{ }
\DeclareUnicodeCharacter{2248}{$\approx$}

\let\Oldsection\section
\renewcommand{\section}{\FloatBarrier\Oldsection}

\let\Oldsubsection\subsection  
\renewcommand{\subsection}{\FloatBarrier\Oldsubsection}

\let\Oldsubsubsection\subsubsection
\renewcommand{\subsubsection}{\FloatBarrier\Oldsubsubsection}

\usepackage[LGRgreek]{mathastext}

\usepackage[margin=1in]{geometry}

\title{\textbf{Structure and equation of state of $Bi_2Sr_2Ca_{n-1}Cu_nO_{2n+4+\delta}$ from x-ray diffraction to megabar pressures}}

\author[a]{Alexander C. Mark}
\author[a]{Muhtar Ahart}
\author[a]{Ravhi Kumar}
\author[b]{Changyong Park}
\author[b]{Yue Meng}
\author[b]{Dmitry Popov}
\author[c]{Liangzi Deng}
\author[c]{Ching-Wu Chu}
\author[a]{Juan Carlos Campuzano}
\author[d]{\\Russell J. Hemley}

\affil[a]{\normalsize Department of Physics, University of Illinois Chicago, Chicago, Illinois 60607, USA}
\affil[b]{\normalsize HPCAT, X-ray Science Division, Argonne National Laboratory, Lemont, Illinois 60439, USA}
\affil[c]{Department of Physics and Texas Center for Superconductivity,}
\affil[ ]{University of Houston, Houston, Texas 77204, USA}
\affil[d]{\normalsize Departments of Physics, Chemistry, and Earth and Environmental Sciences,} 
\affil[ ]{University of Illinois Chicago, Chicago, Illinois 60607, USA}
\date{}

\begin{document}
\maketitle
\pagenumbering{arabic}
\graphicspath{{pics/}}

\begin{abstract}
Pressure is a unique tuning parameter for probing the properties of materials and has been particularly useful for studies of electronic materials such as high-temperature cuprate superconductors. Here we report the effects of quasi-hydrostatic compression produced by a neon pressure-medium on the structures of bismuth-based high $\mathit{T_c}$ cuprate superconductors with the nominal composition $Bi_2Sr_2Ca_{n-1}Cu_nO_{2n+4+\delta}$ (n=1,2,3) up to 155 GPa. The structures of all three compositions obtained by synchrotron X-ray diffraction can be described as pseudo-tetragonal over the entire pressure range studied. We show that previously reported pressure-induced distortions and structural changes arise from the large strains that can be induced in these layered materials by non-hydrostatic stresses. The pressure-volume equations of state (EOS) measured under these quasi-hydrostatic conditions cannot be fit to single phenomenological formulation over the pressure ranges studied, starting below 20 GPa. This intrinsic anomalous compression as well as the sensitivity of $Bi_2Sr_2Ca_{n-1}Cu_nO_{2n+4+\delta}$ to deviatoric stresses provides explanations for the numerous inconsistencies in reported EOS parameters for these materials. We conclude that the anomalous compressional behavior of all three compositions is a manifestation of the changes in electronic properties that are also responsible for the remarkable non-monotonic dependence of $\mathit{T_c}$ with pressure, including the increase in $\mathit{T_c}$ at the highest pressures studied so far for each. Transport and spectroscopic measurements up to megabar pressures are needed to fully characterize and explore still higher possible critical temperatures in these materials.

\end{abstract}

\pagebreak
\setstretch{1}

\begin{refsection}
\section{Introduction}

The nature of unconventional superconductivity in the cuprates is a subject of continued study \cite{chu_hole-doped_2015,agterberg_physics_2020}. Decades of evidence support the hypothesis that superconductivity in these compounds is a quasi-2D phenomena that emerges from the layers of parallel $CuO_2$ planes common to these materials \cite{pickett_electronic_1989,damascelli_angle-resolved_2003,logvenov_high-temperature_2009,bollinger_two-dimensional_2016}. The critical temperature ($\mathit{T_c}$) in the materials can therefore be modified by altering the electronic structure of the $CuO_2$ planes. Though these changes in $\mathit{T_c}$ are typically conducted by variable oxygen doping \cite{kaminski_change_2006}, the application of pressure can also tune $\mathit{T_c}$ in the cuprates in an analogous manner \cite{osada_high-pressure_2000,osada_pressure-induced_2001,neumeier_pressure_1993, almasan_pressure_1992,ambrosch-draxl_pressure-induced_2004,sakakibara_first-principles_2013}. The tendency for the $\mathit{T_c}$-\textit{P} relation to follow a roughly parabolic trajectory that parallels the doping dependence of $\mathit{T_c}$ is reported in nearly all known superconducting cuprates (see reviews \cite{wijngaarden_pressure_1992,wijngaarden_ultra_1993,mark_progress_2022} and references therein). Understanding the nature of unconventional superconductivity, along with the potential to carefully engineer strain conditions to further enhance $\mathit{T_c}$ \cite{choi_3d_2019,zhang_review_2022}, necessitates high pressure structural and equation of state (EOS) studies of these materials.

\begin{figure}[!ht]
\begin{centering}
\includegraphics[width=0.75\textwidth]{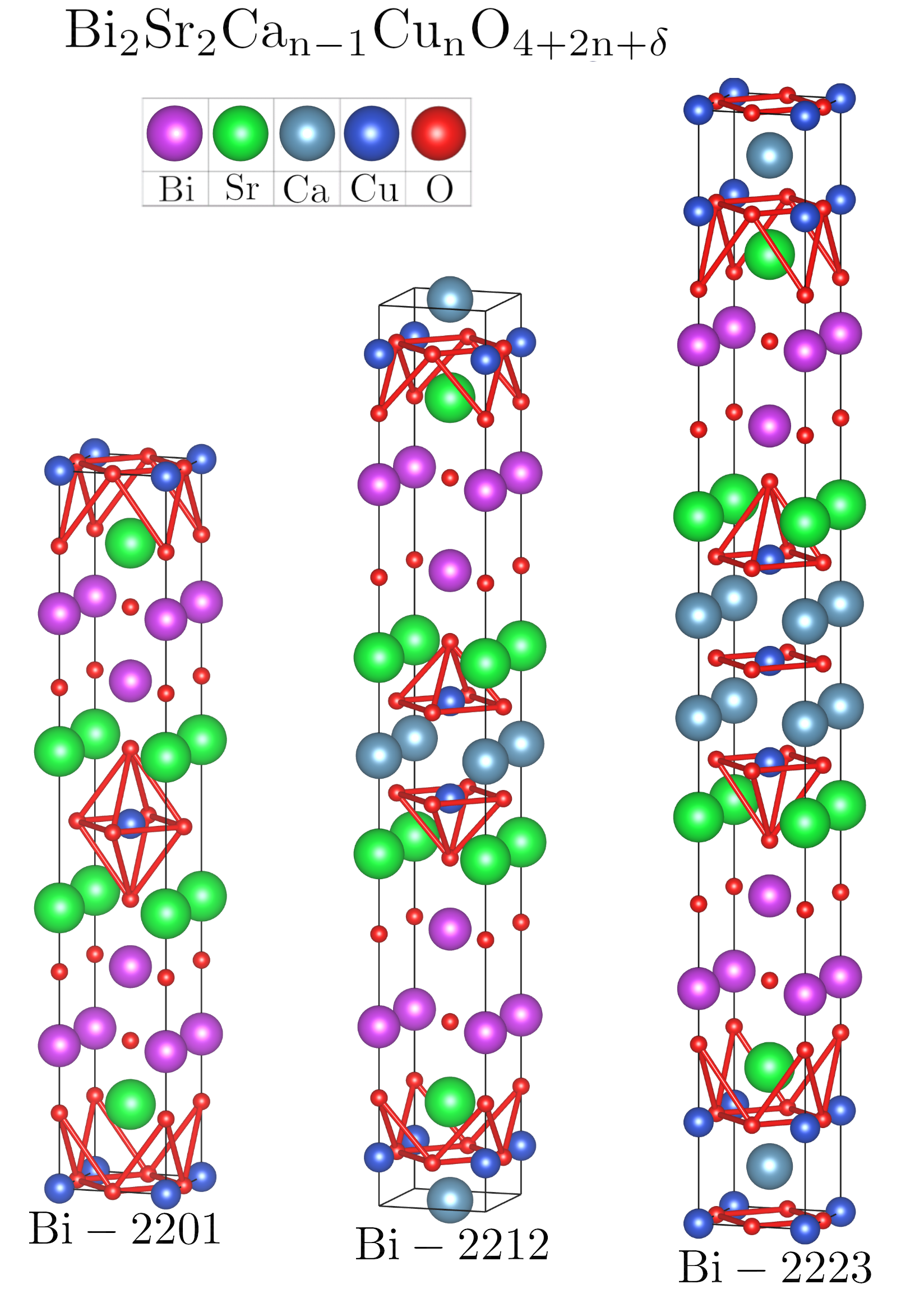}
\caption{Unit cells of Bi-2201, Bi-2212, and Bi-2223 presented in a tetragonal (\textit{I}4-\textit{mmm}) representation. Structural parameters for Bi-2201 from Matheis et al. \cite{matheis_crystal_1990} and for Bi-2212 and Bi-2223 from Wesche et al. \cite{wesche_high-temperature_2017}.}
\label{BSCCO-struct}
\end{centering}
\end{figure}

The effect of pressure on the bismuth based cuprate superconductors [$Bi_2Sr_2Ca_{n-1}Cu_nO_{2n+4+\delta}$ (n=1-3) (BSCCO)] are particularly intriguing. In all three compounds $\mathit{T_c}$ is observed to first increase then decrease on further compression \cite{chen_enhancement_2010,deng_higher_2019,zhou_quantum_2022,matsumoto_crystal_2021,chen_enhancement_2010}. 
The behavior of $\mathit{T_c}$ upon continued pressurization is dependent on both compression environment and oxygen doping; in some experiments $\mathit{T_c}$ continues to decrease and superconductivity is eventually destroyed \cite{zhou_quantum_2022}, while in other experiments as pressure is increased $\mathit{T_c}$ breaks away from the dome-like trend and increases to the maximum pressures reported for each compound. Chen et al. \cite{chen_enhancement_2010} discovered this `up-down-up' trajectory with pressure in $Bi_2Sr_2Ca_2Cu_3O_{10+\delta}$ (Bi-2223) with the second increase beginning at 25 GPa. Deng et al. \cite{deng_higher_2019} later found this behavior in $Bi_2Sr_2CuO_{6+\delta}$ (Bi-2201) and in $Bi_2Sr_2CaCu_2O_{8+\delta}$ (Bi-2212) where the second rise in $\mathit{T_c}$ begins at 45 GPa and 40 GPa, respectively. Similarly non-monotonic $\mathit{T_c}$ pressure dependencies have not been reported to date in any other cuprate superconductors. All compounds are reported to maintain a layered perovskite structure (Fig. \ref{BSCCO-struct}) to at least 50 GPa \cite{hervieu_electron_1988,shamrai_crystal_2013, antipov_structural_2008}, so these changes in $\mathit{T_c}$ can be directly correlated with compression of the prototype structure of each. In contrast, other cuprate superconductors undergo pressure-induced structural transitions \cite{nakayama_collapse_2014,ludwig_x-ray_1990,nelmes_crystal_1990,takagi_disappearance_1992} including hysteresis effects \cite{looney_influence_1998,sadewasser_relaxation_1999} which can influence or destroy superconductivity.

High pressure structural data have been reported for Bi-2212 \cite{olsen_high_1991,gavarri_anisotropic_1990,zhang_pressure-induced_2013,zhou_quantum_2022}, but the results are conflicting. Olsen et al. \cite{olsen_high_1991} studied compression of a mixed phase sample of Bi-2212 and Bi-2223 to 50 GPa using energy-dispersive x-ray diffraction (XRD). The group observed a stiffening of the \textit{c} parameter near 20 GPa, followed by a drastic decrease in \textit{c} between 35 and 40 GPa. Deng et al. \cite{deng_higher_2019} noted that this decrease occurs near the pressure of the rise in $\mathit{T_c}$ in Bi-2212, and suggest that this structural change arises from a pressure-induced Lifshitz transition. Zhang et al. \cite{zhang_pressure-induced_2013} report significantly different pressure-induced distortions in Bi-2212 on compression,
specifically a transition from the orthorhombic (pseudo-tetragonal) structure to a `collapsed orthorhombic' structure near 20 GPa. Zhang et al. \cite{zhang_pressure-induced_2013} observed a pressure-induced stiffening in \textit{c} near 20 GPa, but did not reach the pressures where the collapse in \textit{c} had been reported \cite{olsen_high_1991}. In addition to these structural discrepancies, considerable differences in equation of state (EOS) parameters for BSCCO have been reported. Results obtained using different techniques produce different results; for example, fitting XRD data for Bi-2212 gives a bulk modulus ($\mathit{K_{0}}$) of 60-130 GPa \cite{olsen_high_1991,gavarri_anisotropic_1990,zhang_pressure-induced_2013,zhou_quantum_2022} whereas direct ambient pressure ultrasonic measurements give a $\mathit{K_{0}}$ of 15-40 GPa \cite{solunke_effect_2007,fanggao_ultrasonic_1991-1,seetawan_synthesis_2008}.

Here we report a detailed structural and EOS study of $Bi_2Sr_2Ca_{n-1}Cu_nO_{2n+4+\delta}$  n=1,2,3 compounds (Bi-2201, Bi-2212, Bi-2223, respectively) under quasi-hydrostatic compression up to megabar pressures. Quasi-hydrostatic conditions were achieved by pressurizing samples with a neon medium \cite{rivers_compresgsecars_2008}, which remains significantly more hydrostatic under pressure than the media used in the past for BSCCO \cite{klotz_hydrostatic_2009}. This approach allows us to measure the compressive properties of BSCCO under considerably more hydrostatic stresses and extend the pressure range of reported \textit{P}-\textit{V} data. Experiments were also performed using no pressure transmitting medium in order to compare with our quasi-hydrostatic and previous non-hydrostatic results. The data reveal all three BSCCO compounds to be highly susceptible to deviatoric stress. We now attribute many of the previously reported pressure-induced structural distortions to artifacts induced by non-hydrostatic stress conditions. Additionally, even when compressed quasi-hydrostatically, all three materials exhibit anomalous compression with \textit{P}-\textit{V} relations not being well described by a single standard EOS formulation.
\section{Methods}
 
Samples were loaded in symmetric diamond anvil cells (DACs) with diamond culet sizes ranging from 300 - 600  $\mu m$. The single crystal Bi-2201 and Bi-2212 samples were optimally doped and synthesized at the Houston Center for Superconductivity. Sintered pellets of Bi-2212 + Bi-2223 were purchased from Quantum Levitation. Composition and ambient pressure unit cell volumes ($\mathit{V_0}$) were determined using ambient pressure XRD utilizing a Bruker D8 Advance diffractometer with a Cu source. The volume ratio of the Bi-2223/Bi-2212 samples were estimated to be 80\%/20\%. All samples were ground into a fine powder using an agate mortar and pestle. Tungsten and rhenium gaskets were pre-indented to 20 GPa and holes of 50 - 150 $\mu m$ in diameter were laser drilled to form the sample chambers \cite{hrubiak_laser_2015}. Samples were loaded in DACs along with neon as a pressure transmitting medium (for our quasi-hydrostatic runs) at GSECARS, Sector 13, APS, ANL \cite{rivers_compresgsecars_2008}. Neon was chosen for these experiments because, even after pressure-induced freezing at 4.8 GPa \cite{finger_structure_1981} at room temperature, the material remains a weak solid to megabar pressures \cite{hemley_x-ray_1989} and is  considerably more hydrostatic than the pressure transmitting media used in the past for BSCCO XRD studies \cite{klotz_hydrostatic_2009,olsen_high_1991,gavarri_anisotropic_1990,zhang_pressure-induced_2013,zhou_quantum_2022}. XRD was measured at the 16-IDB and 16-BMD beamlines at HPCAT, Sector 16, APS, ANL \cite{park_new_2015}. Pressures were determined from lattice parameters obtained by XRD of gold flakes placed within the sample chamber using the EOS of Anderson et al. \cite{anderson_anharmonicity_1989}. The pressures were consistent with those determined by ruby fluorescence measurements using the Xu et al. \cite{xu_high-pressure_1986} calibration.  


BSCCO compounds are reported to maintain orthorhombic symmetry similar to the ambient pressure \textit{Ammm} structure \cite{hervieu_electron_1988,shamrai_crystal_2013} over the pressure ranges studied to date \cite{olsen_high_1991, gavarri_anisotropic_1990, zhang_pressure-induced_2013, zhou_quantum_2022}. At ambient pressures the BSCCO unit cells can be described as pseudo-tetragonal (space group \textit{I}4/\textit{mmm}) with $\mathit{a} - \mathit{b} \leq 0.1 \; \text{\AA}$ \cite{olsen_high_1991,gavarri_anisotropic_1990}. This representation appears to be valid at high pressure, and is adopted in the Le Bail refinements performed in this work using JANA 2006 \cite{petricek_crystallographic_2014} to determine lattice parameters. Patterns were fit to a unit cell with \textit{a}$\approx 3.8$ \AA \; and a compound dependent \textit{c} axis length. In order to directly compare with past results, in which the \textit{ab} plane of the primitive unit cell was rotated $45^o$ in order to place the cell boundary along the Cu-Cu bond resulting in an ambient pressure $\mathit{a}\approx 5.4$ \AA ,  the \textit{a} parameters from our refinements were multiplied by $\sqrt{2}$. Though BSCCO compounds are known to exhibit both commensurate \cite{ariosa_periodic_2001,timofeev_growth_1998} and incommensurate \cite{zandbergen_electron_1988,timofeev_growth_1998,poccia_spatially_2020} modulations along the \textit{b} and \textit{c} axes respectively, diffraction peaks from these modulations are low intensity and typically not observed in XRD experiments of a comparable sensitivity to the probes used in this work \cite{tarascon_preparation_1988}. For the three materials, all prominent peaks in the collected diffractograms were indexed to Bragg reflections with no peaks attributed to supermodulation of the structures. The zero pressure bulk modulus ($\mathit{K_0}$) and its pressure derivative ($\mathit{K_{0}^{'}}=\frac{d\mathit{K_0}}{d\mathit{P}} $) were determined by fitting the \textit{P}-\textit{V} data to a Vinet EOS \cite{vinet_temperature_1987} given as

\begin{equation}
\label{eqn:vinet}
\mathit{P} = 3\mathit{K_0} \bigg( \frac{1-\eta}{\eta^2}\bigg)\exp\bigg[{\frac{3}{2}(\mathit{K'_0} -1)(1-\eta)}\bigg],
\end{equation}

\noindent using a least squares method, where $\eta$ is the strain ($\sqrt[3]{\mathit{V} / \mathit{V_o}}$).

\section{Results}

\subsection{Bi-2201}

\begin{figure}[!ht]
\begin{centering}
\includegraphics[width=0.9\textwidth]{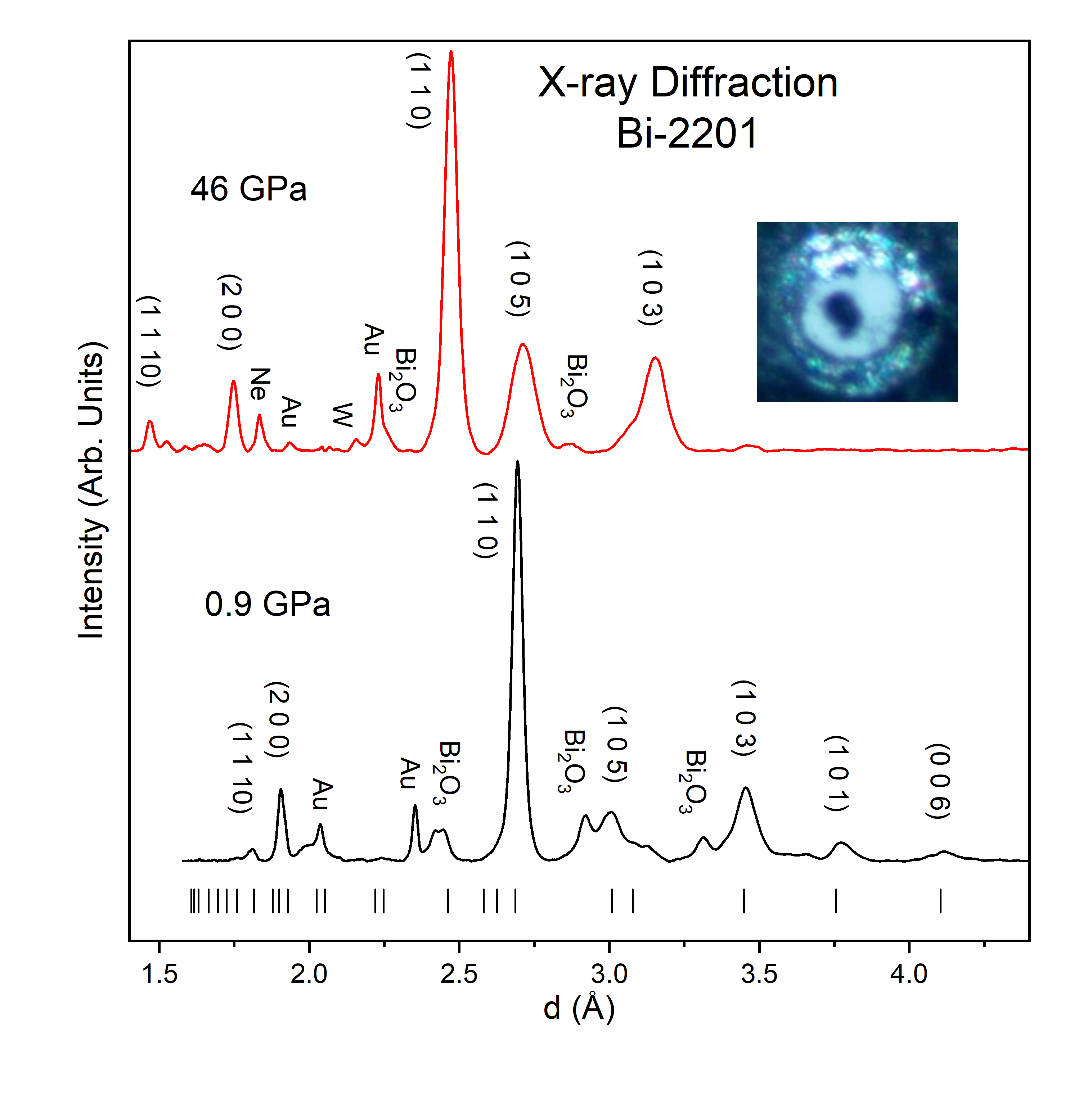}
\caption{Representative radially integrated x-ray diffraction patterns and peak indexing of single phase Bi-2201 at 0.9 GPa (black) and 46 GPa (red) ($\lambda=$ 0.4959 \AA). Tick marks indicate predicted Bragg peaks for Bi-2201 at 0.9 GPa. Bi-2201 was fit to a tetragonal \textit{I}4/\textit{mmm} symmetry unit cell with ambient pressure parameters \textit{a} = 3.81(1) $\text{\text{\AA}}$ and c = 24.6(1) $\text{\text{\AA}}$. Peaks from unreacted $Bi_2O_3$ from synthesis are identified in the pattern. Inset is a reflected light micrograph of the sample loaded in the DAC at 0.9 GPa with a 150 $\mu$m sample chamber diameter surrounded by the pressure-transmitting medium.} 
\label{2201-index}
\end{centering}
\end{figure}

\begin{figure}[!ht]
\begin{centering}
\includegraphics[width=\textwidth]{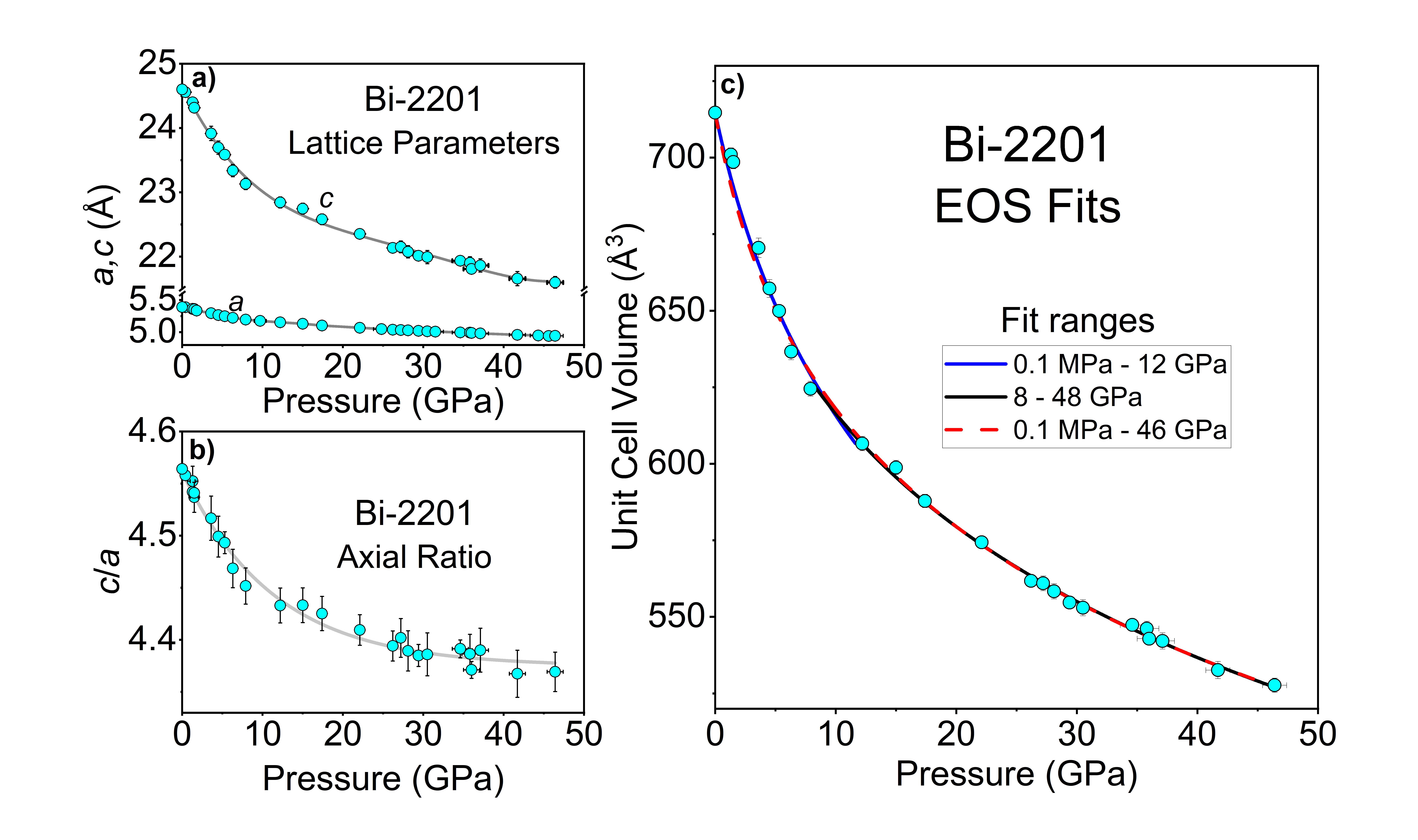}
\caption{Pressure dependence of the structural parameters of Bi-2201. \textbf{(a)} \textit{a} (red) and \textit{c} (blue) parameters of Bi-2201. \textbf{(b)}\textit{c}/\textit{a} ratio of Bi-2212, showing a change in the observed trend between 8-10 GPa. Lines inserted as guides to the eye. \textbf{(c)} Unit cell volume with Vinet EOS fit to the data over three pressure ranges, 0.1 MPa - 48 GPa [$\mathit{V_0}$=714.67 \AA$^3$, $\mathit{K_0}$=32.0(1.3) GPa, $\mathit{K_{0}^{'}}$=10.7(3)], 0.1 MPa - 12 GPa [$\mathit{V_0}$=714.67 \AA$^3$, $\mathit{K_0}$=38.1(5.1) GPa, $\mathit{K_{0}^{'}}$=7.8(2.0)], and 8 - 48 GPa [$\mathit{V_0}$=696.6(4.0) \AA$^3$, $\mathit{K_0}$=45.8(3.4) GPa, $\mathit{K_{0}^{'}}$=9.6(4)]. For pressure ranges including zero pressure, $\mathit{V_0}$ was fixed to 714.67 \AA$^3$ which we determined using ambient pressure XRD. All error bars represent 1 sigma uncertainty.} 
\label{2201-measured}
\end{centering}
\end{figure}

Polycrystalline XRD of Bi-2201 was measured to 46 GPa (Fig. \ref{2201-index}). Given the medium resolution of the measurements, the data can be fit with a pseudo-tetragonal structure (\textit{I}4/\textit{mmm} symmetry) with ambient pressure unit cell parameters [a=5.39(1) \AA, c=24.6(1) \AA] that are in agreement with those reported previously (see Supplementary Materials Table S\ref{table:2201-ambient}) \cite{pryanichnikov_nonmonotonic_2012, khasanova_bi-2201_1995, mironov_new_2016}. Several peaks are attributed to a small amount of $Bi_2O_3$, a common precursor material in BSCCO synthesis \cite{khasanova_bi-2201_1995} still present in the sample; pressure dependence of these peaks are consistent with previously reported data \cite{pereira_structural_2013,fredenburg_high-pressure_2011}. The \textit{a} and \textit{c} lattice parameters of Bi-2201 are found to decrease monotonically with pressure [Fig. \ref{2201-measured}(a)]; \textit{a} decreases smoothly with pressure whereas \textit{c} decreases but becomes less compressible between 8 and 10 GPa. This change in compressibility is also clearly evident in the pressure dependence of the \textit{c}/\textit{a} ratio [Fig. \ref{2201-measured}(b)].

Fitting the entire dataset to a single Vinet EOS with a fixed $\mathit{V_0}$ determined using ambient pressure XRD resulted in unacceptably large residuals, especially below 5 GPa where the measured unit cell volumes at a given pressure are up to 10 \AA$^3$ higher than the best-fit EOS. Above 10 GPa the residuals are much smaller. We are able to greatly improve the fits by splitting the data into two regions, above and below the point of stiffening in \textit{c} at 8 GPa [Fig. \ref{2201-measured}(c)]. In the high pressure fit all three parameters were allowed to vary, while the low pressure fits $\mathit{V_0}$ was fixed to 714.67 \AA$^3$ which we measured at ambient pressure. 

\subsection{Bi-2212}

\begin{figure}[!ht]
\begin{centering}
\includegraphics[width=0.9\textwidth]{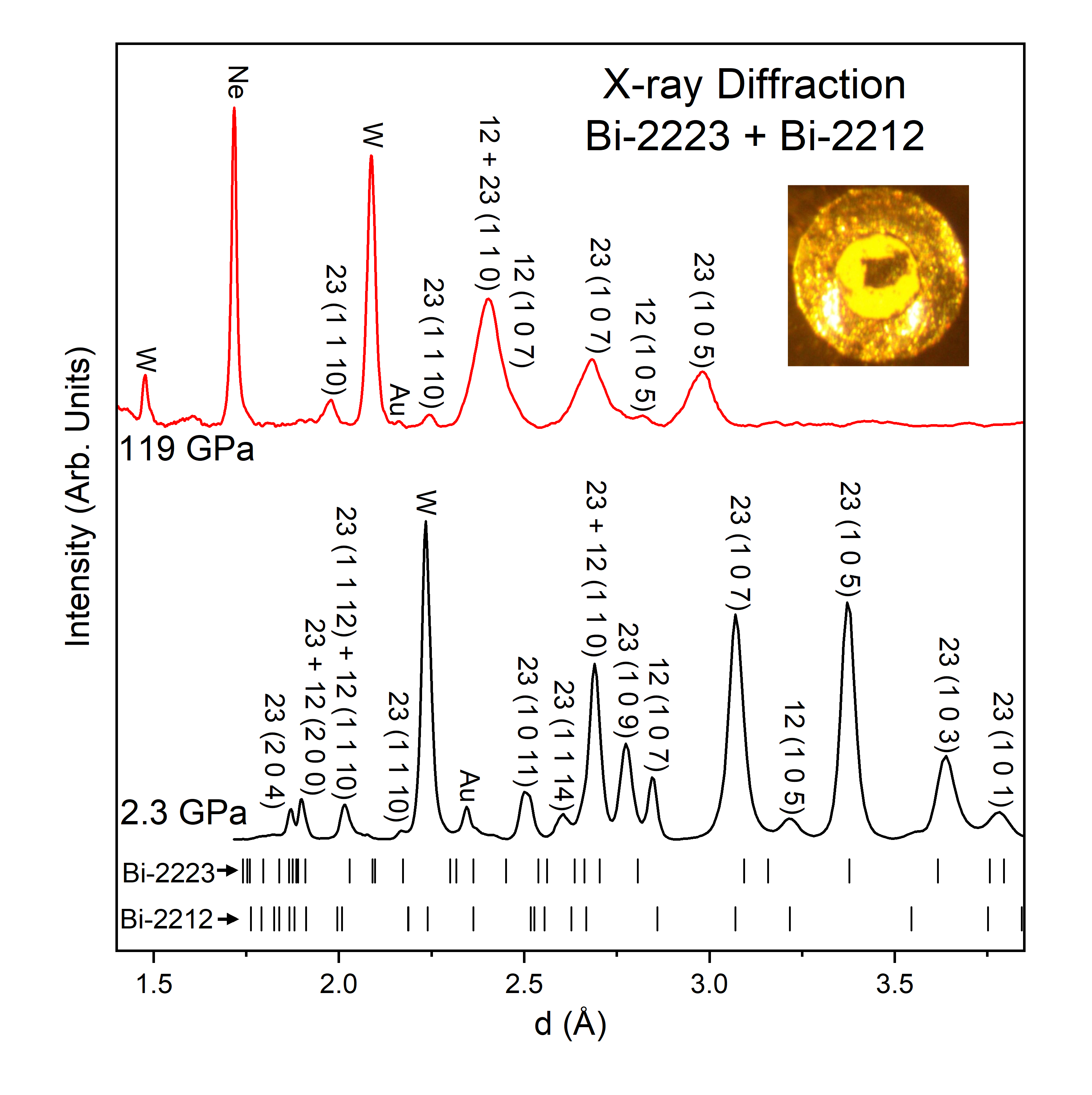}
\caption{Representative x-ray diffraction pattern and peak indexing of mixed phase Bi-2212 + Bi-2223 at 2.3 GPa (black, $\lambda$=0.406626 \AA) and 119 GPa (red, $\lambda$=0.4133 \AA). Both are indexed using tetragonal \textit{I}4/\textit{mmm} symmetry. Tick marks indicate predicted Bragg peaks of Bi-2223 (black) and Bi-2212 (red) at 2.3 GPa. Diffraction from the pressure transmitting medium Ne (at pressures above pressure-induced freezing) and W from the the gasket are identified. Inset is a reflected light micrograph of mixed phase Bi-2212 and Bi-2223 at 2.3 GPa in the DAC with a 150 $\mu$m sample chamber surrounded by the pressure-transmitting medium.} 
\label{mixed-index}
\end{centering}
\end{figure}

\begin{figure}[!ht]
\begin{centering}
\includegraphics[width=\textwidth]{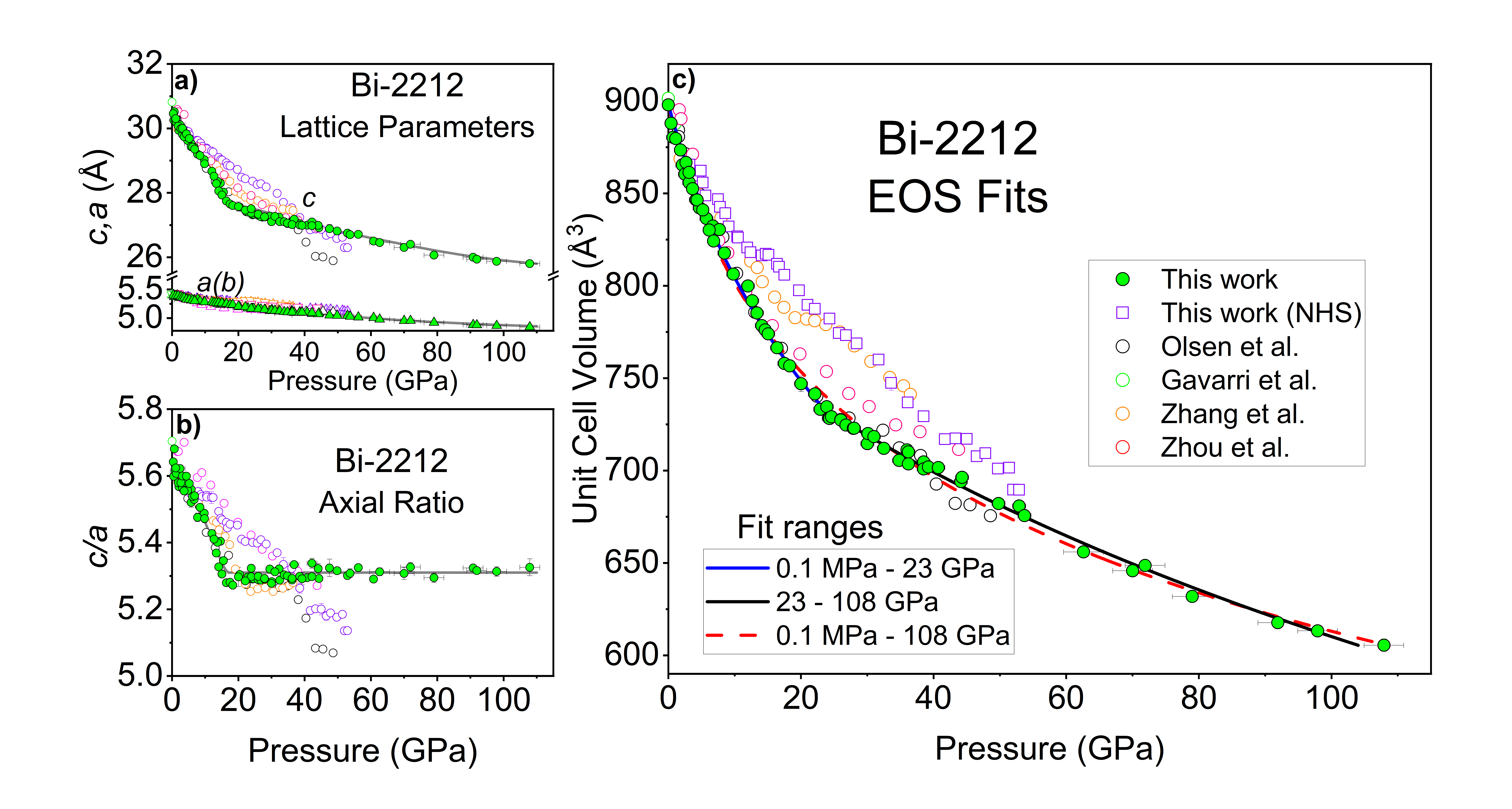}
\caption{Pressure dependence of the structural parameters of Bi-2212 measured under quasihydrostatic conditions (this work, green solid circles) and non-hydrostatic conditions (NHS; this work, open purple squares) and previous results, Olsen et al. \cite{olsen_high_1991}, Gavarri et al. \cite{gavarri_anisotropic_1990}, Zhang et al. \cite{zhang_pressure-induced_2013}, and Zhou et al. \cite{zhou_quantum_2022}. \textbf{(a)}\textit{a} (triangles), \textit{b} (inverted triangles), and \textit{c} (circles) lattice parameters. Lines inserted as guides to the eye. \textbf{(b)}\textit{c}/textit{a} axial ratio, showing a change in compression mechanism at 23 GPa, coinciding with the stiffening of \textit{c}. \textbf{(c)} Pressure dependence of the unit cell volume and Vinet EOS fit to the data over three pressure ranges, 0.1 MPa - 108 GPa [$\mathit{V_0}$=898.0 \AA$^3$, $\mathit{K_0}$=53.9(1.6) GPa, $\mathit{K_{0}^{'}}$=8.7(2)] 0.1 MPa - 23 GPa [$\mathit{V_0}$=898.0 \AA$^3$, $\mathit{K_0}$=67.7(2.3) GPa, $\mathit{K_{0}^{'}}$=4.9(3)], and 23 - 108 GPa [$\mathit{V_0}$=793.9(4.6) \AA$^3$, $\mathit{K_0}$=262.8(22.4) GPa, $\mathit{K_{0}^{'}}$=2.9(4)]. For pressure ranges containing ambient pressure (0.1 MPa), $\mathit{V_0}$ was fixed to 898.0 \AA$^3$, a value obtained by ambient pressure XRD. Low pressure and high pressure fits produce $\mathit{K_{0}}$ and $\mathit{K_{0}^{'}}$ values in agreement with each other over similar pressure ranges. For all plots error bars represent 1 sigma uncertainty.} 

\label{2212-measured}
\end{centering}
\end{figure}

Powdered XRD patterns for single phase Bi-2212 were measured to 61 GPa and mixed phase Bi-2212 + Bi-2223 up to 108 GPa over five experimental runs. Similar to Bi-2201, the patterns were described with pseudo-tetragonal \textit{I}4/\textit{mmm} symmetry [$\mathit{a}=5.404(1)\text{\AA}, \mathit{c}=30.750(5)\text{\AA}$] with ambient pressure lattice parameters in agreement with previous reports (see Supplementary Materials Table S\ref{table:2212-ambient}) \cite{hervieu_electron_1988,petricek_x-ray_1990,gao_combined_1993,olsen_high_1991,gavarri_anisotropic_1990,babaei_pour_determination_2005,tarascon_crystal_1988}. Representative XRD patterns are presented in Fig. \ref{mixed-index}. As is observed in Bi-2201, \textit{a} decreases monotonically, while \textit{c} exhibits a stiffening at 18 - 20 GPa [Fig. \ref{2212-measured}(a, b)]. Our \textit{P}-\textit{V} results [Fig. \ref{2212-measured}(c)] overlap with previously reported data up to 15 GPa \cite{olsen_high_1991,zhang_attenuation_2013,zhou_quantum_2022}, including the reported onset of stiffening in \textit{c} \cite{olsen_high_1991, zhang_pressure-induced_2013}. Above 15 GPa, however, our quasi-hydrostatic data diverge from previously reported results. Olsen et al. \cite{olsen_high_1991} observed a decrease of nearly 7\% in the \textit{c}/\textit{a} axial ratio from 38-42 GPa, whereas our data indicate that the axial ratio saturates at pressures above 20 GPa and remains roughly constant to the upper pressure limit of our measurement. In contrast to Zhang et al. \cite{zhang_pressure-induced_2013} we find no evidence for a collapsed orthorhombic phase under pressure, with the pseudo-tetragonal unit cell describing the structure over the pressure range of the experiment. The pressure response of \textit{a} and \textit{b} remains close over the measured range, and no growth in \textit{a} or \textit{b} is observed. In addition, above 15 GPa our unit cell volumes are consistently lower than those reported by Zhou et al. \cite{zhou_quantum_2022}.

To test the impact of hydrostaticity on the compression properties of Bi-2212, we performed XRD measurements with no pressure transmitting medium by directly compressing the sample between two diamonds. This non-hydrostatic compression exhibited several anomalies in the \textit{a} and \textit{c} parameters similar to those observed in previous experiments. Most notably we observe a pressure-induced increase in \textit{a} similar to that observed by Zhang et al. \cite{zhang_pressure-induced_2013}, and a series of collapses in \textit{c} similar to those observed by Olsen et al. \cite{olsen_high_1991}. A quantitative comparison between the features apparent in our non-hydrostatic experiment and those present in previously reported data is not possible because the degree of non-hydrostaticity depends on unreported details of the previous experiments. 

When fitting all quasi-hydrostatic data (0.1 MPa - 110 GPa) to a single EOS we obtain a $\mathit{K_0}$ of 38.3(1.3) GPa, with $\mathit{V_0}$ = 898.0(1)\AA$^3$, and $\mathit{K_{0}^{'}}$ = 8.8(3). This $\mathit{K_0}$ value is significantly lower than those reported in the literature for data based on analysis of high-pressure XRD data \cite{tajima_pressure-effect_1989,yoneda_pressure-effect_1990,olsen_high_1991,zhang_pressure-induced_2013,zhou_quantum_2022}, and slightly higher than those reported using ultrasonic measurements \cite{solunke_effect_2007,fanggao_ultrasonic_1991-1}. These discrepancies motivated us to fit the data over multiple different pressure regimes as was done for Bi-2201. For ranges that contain ambient pressure, the data was fit to a Vinet EOS with $\mathit{V_0}$ fixed at 898.4 \AA$^3$ as determined from ambient pressure XRD [Fig. \ref{2212-measured}(c)] with $\mathit{K_0}$ and $\mathit{K_{0}^{'}}$ allowed to vary. Above 23 GPa, the data was fit to a Vinet EOS with all parameters allowed to vary. Splitting \textit{P}-\textit{V} data in this way results in greatly improved fitting of the data over the entire pressure range.

\subsection{Bi-2223}

\begin{figure}[!ht]
\begin{centering}
\includegraphics[width=\textwidth]{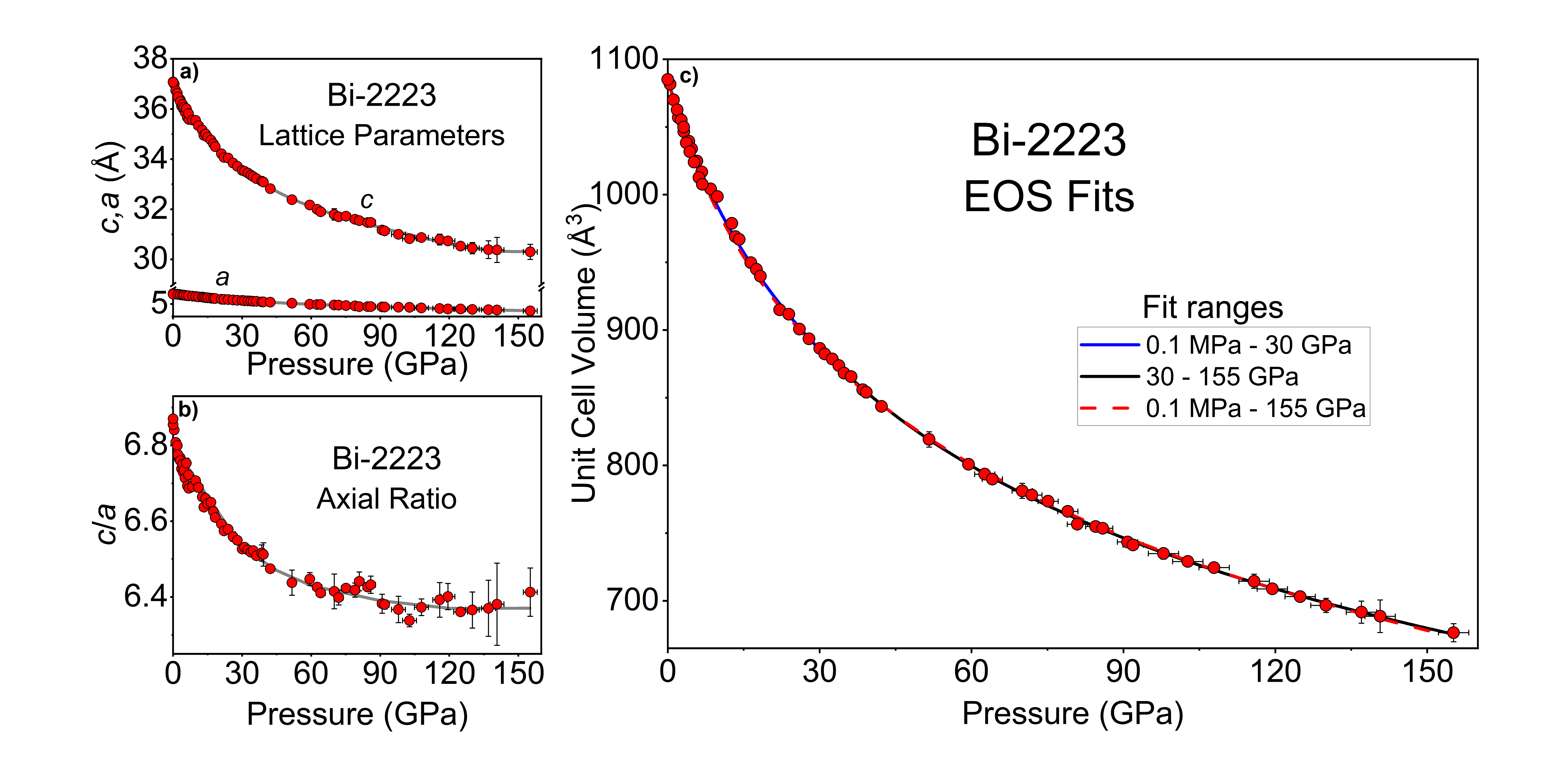}
\caption{Pressure dependence of the structural parameters of Bi-2223. \textbf{(a)} \textit{a} and \textit{c} parameters. Lines inserted as guides to the eye. \textbf{(b)} \textit{c}/\textit{a} ratio for Bi-2223. \textbf{(c)}Unit cell volume with Vinet EOS fit to the data over three pressure ranges, 0.1 MPa - 155 GPa [$\mathit{V_0}$=1085.01 \AA$^3$, $\mathit{K_0}$=77.7(9) GPa, $\mathit{K_{0}^{'}}$=6.4(1)], 0.1 MPa - 30 GPa[$\mathit{V_0}$=1085.01 \AA$^3$, $\mathit{K_0}$=84.9(2) GPa, $\mathit{K_{0}^{'}}$=5.6(4)], and 30-155 GPa [$\mathit{V_0}$=1179(36) \AA$^3$, $\mathit{K_0}$=35.5(9) GPa, $\mathit{K_{0}^{'}}$=7.9(5)]. For pressure ranges including ambient pressure (0.1 MPa), $\mathit{V_0}$ was fixed to 1085.01 \AA$^3$, a value obtained from ambient pressure XRD. All error bars represent 1 sigma uncertainty.}
\label{2223-measured}
\end{centering}
\end{figure}

XRD of from the mixed phase Bi-2212 + Bi-2223 samples provided structural information of Bi-2223 to 155 GPa (Fig. \ref{2223-measured}). Similar to the other two materials, the XRD patterns are well described by a pseudo-tetragonal \textit{I}4/\textit{mmm} structure [$\mathit{a}=5.409(5)\text{\AA}, \mathit{c}=37.080(6)\text{\AA}$] with ambient pressure lattice parameters in agreement with previously reported results (see Supplementary Materials Table S\ref{table:2223-ambient}) \cite{shamray_crystal_2009,maljuk_floating_2016,giannini_growth_2005}.

As with Bi-2212, $\mathit{K_{0}^{'}}$ is reported to be significantly larger when measured using ultrasonic probes, with Fanggao et al. \cite{fanggao_ultrasonic_1991-1} reporting values between 39-59 whereas our XRD measurements indicate a value for $\mathit{K_{0}^{'}}$ closer to 4.1(3). Our high pressure fit indicates that $\mathit{K_0}$ decreases to 34.3 GPa with $\mathit{K_{0}^{'}} \approx$ 8.1(5). Like Bi-2201 and Bi-2212, no change in the compression of the \textit{a} axis was apparent, and a slight kink in \textit{c} was observed at 30 GPa which is more obvious in the pressure dependence of the \textit{c}/\textit{a} ratio (Fig. \ref{2223-measured}). As with Bi-2201 and Bi-2212, this feature appears to be due to a stiffening in the \textit{c} axis.

Similar to the the other two compounds, the data could not be fit to a single EOS over the entire pressure range studied. Attempting to fit the data to a single EOS resulted in unacceptably large residuals, especially below 30 GPa [Fig. \ref{2223-measured}(c)]. Therefore, the data was split into two regimes at 30 GPa and fit to two separate EOS. As determined previously for the other two materials, splitting the fits resulted in $\mathit{V_0}$, $\mathit{K_{0}}$, and $\mathit{K_{0}^{'}}$ parameters with significantly smaller residuals and uncertainties. Select fits are presented in Fig. \ref{2223-measured} (c). Fits that contained the volume at ambient pressure used a fixed $\mathit{V_o}$ = 1085.01 \AA$^3$ which we determined using ambient pressure XRD.




\section{Discussion}

\subsection{Equation of State Fits}

As discussed previously, in all three materials fitting the data to a single phenomenological EOS over the entire measured pressure range produces large residuals, especially at low pressures (Figs \ref{2201-measured}-\ref{2223-measured}). This result is independent of which condensed phase EOS function is used (e.g. Vinet versus Birch-Murnaghan \cite{birch_finite_1947}). Fitting \textit{P}-\textit{V} data in different regimes produces much improved fits. In the lower pressure range, the EOS fit parameters are expected to match those obtained from other methods, such as the ultrasonic determination of the ambient pressure bulk modulus for a sample of the same density and suitably corrected from adiabatic to isothermal environments (e.g. using the methods outlined by Overton \cite{overton_relation_1962}). Unexpectedly, reported values for $\mathit{K_0}$ measured using other techniques, such as ultrasonic measurements, give significantly lower values of $\mathit{K_0}$ than those determined based on fits to high pressure XRD data \cite{olsen_high_1991,gavarri_anisotropic_1990,zhang_pressure-induced_2013, zhou_quantum_2022}. We note that some disagreement in the measured $\mathit{K_0}$ may be attributed to porosity and lattice defects in as-grown cuprate superconductors \cite{al-kheffaji_elastic_1989,cankurtaran_bulk_1989} which are known to be particularly prevalent in as grown Bi-2201 \cite{jin_growth_2000,gorina_dependence_2003} and Bi-2223 \cite{shamray_crystal_2009}. These defects most likely account for the large spread in measured adiabatic $\mathit{K_0}$ values apparent in Bi-2212 (see Fig. \ref{2212-K0} and Supplementary Material Tables S\ref{table:2201-elastic}, S\ref{table:2212-elastic}, S\ref{table:2223-elastic}). Differences in the degree of strain throughout the samples might result in the measured $\mathit{K_0}$ falling between the Voigt \cite{voigt_ueber_1889} and Reuss \cite{reuss_berechnung_1929} bounds, but it is not expected that this spread would be large enough to account for the 2-3 times difference that is observed. 


In Bi-2201 the low pressure (below 8 GPa) fit produced a $\mathit{K_0}$ of 54.5 GPa a value significantly larger than those determined using ultrasonic probes, which tend to measure $\mathit{K_0}$ to be between 15 - 39 GPa \cite{seetawan_synthesis_2008,dominec_elastic_1992}. The fit also resulted in a $\mathit{K_{0}^{'}}$ value of 0.57, which is significantly lower than most materials that are well described with a phenomenological EOS that typically have a $\mathit{K_{0}^{'}} \approx 4$  \cite{cohen_accuracy_2000}. The high pressure fit (above 8 GPa) results in a $\mathit{K_0}$ of 45.8 and a $\mathit{K_{0}^{'}}$ of 9.6. The discrepancy in these values provide further evidence of a change in compression mechanism in Bi-2201 above 8 GPa. These $\mathit{K_0}$ values are also significantly larger than what is has been reported in ultrasonic studies \cite{seetawan_synthesis_2008,dominec_elastic_1992} (Supplemental Materials Table S\ref{table:2201-elastic}). We attribute this probe dependent discrepancy in measured bulk properties to an anomalous compression mechanism in Bi-2201 even below the prominent change in compression at 8 GPa; i.e., Bi-2201 is not well described by a single standard phenomenological EOS even at low pressures. 


Similar to Bi-2201, the $\mathit{K_0}$ values of Bi-2212 measured using ultrasonic probes tend to be significantly lower than those determined via fitting an EOS to \textit{P}-\textit{V} data. The XRD data were fit to two separate compression regions with pressure cutoffs determined by the stiffening in \textit{c} at $\sim$23 GPa as outlined above. For the low pressure fit we find $\mathit{K_0} \approx$ 70.5 GPa, which is comparable to values measured by Olsen et al. \cite{olsen_high_1991}, Yoneda et al. \cite{yoneda_pressure-effect_1990} and Tajima et al. \cite{tajima_pressure-effect_1989} using similar diffraction techniques. Fitting the data gives a $\mathit{K_{0}^{'}} \approx$ 4.9(3), comparable to the fixed value of 4 used in previous studies. On the other hand, ultrasonic measurements tend to produce $\mathit{K_0}$ values between 10-27 GPa \cite{seetawan_synthesis_2008}. For the high pressure fits, the best fit $\mathit{K_0}$ drastically increases to 263(22) GPa and $\mathit{K_{0}^{'}}$ decreases from 4.9(3) to 2.9(4), consistent with a pressure-induced stiffening. 

Our data for Bi-2212 below 20 GPa are in excellent agreement with previously reported \textit{P}-\textit{V} results, (Fig. \ref{2212-measured}). Unlike previous experiments, above 20 GPa we do not observe a sudden collapse in the \textit{c} \cite{olsen_high_1991} or a growth in the \textit{a} \cite{zhang_pressure-induced_2013} parameters. We attribute this difference to our use of Ne as a pressure transmitting medium in contrast to the 4:1 methanol/ethanol mixture used in earlier studies, a mixture which is known to become highly non-hydrostatic above 10 GPa \cite{klotz_hydrostatic_2009}. We therefore attribute both phenomena to non-hydrostatic compression from the sample environment rather than new structural phases formed through hydrostatic compression.

\begin{figure}[!ht]
\begin{centering}
\includegraphics[width=0.5\textwidth]{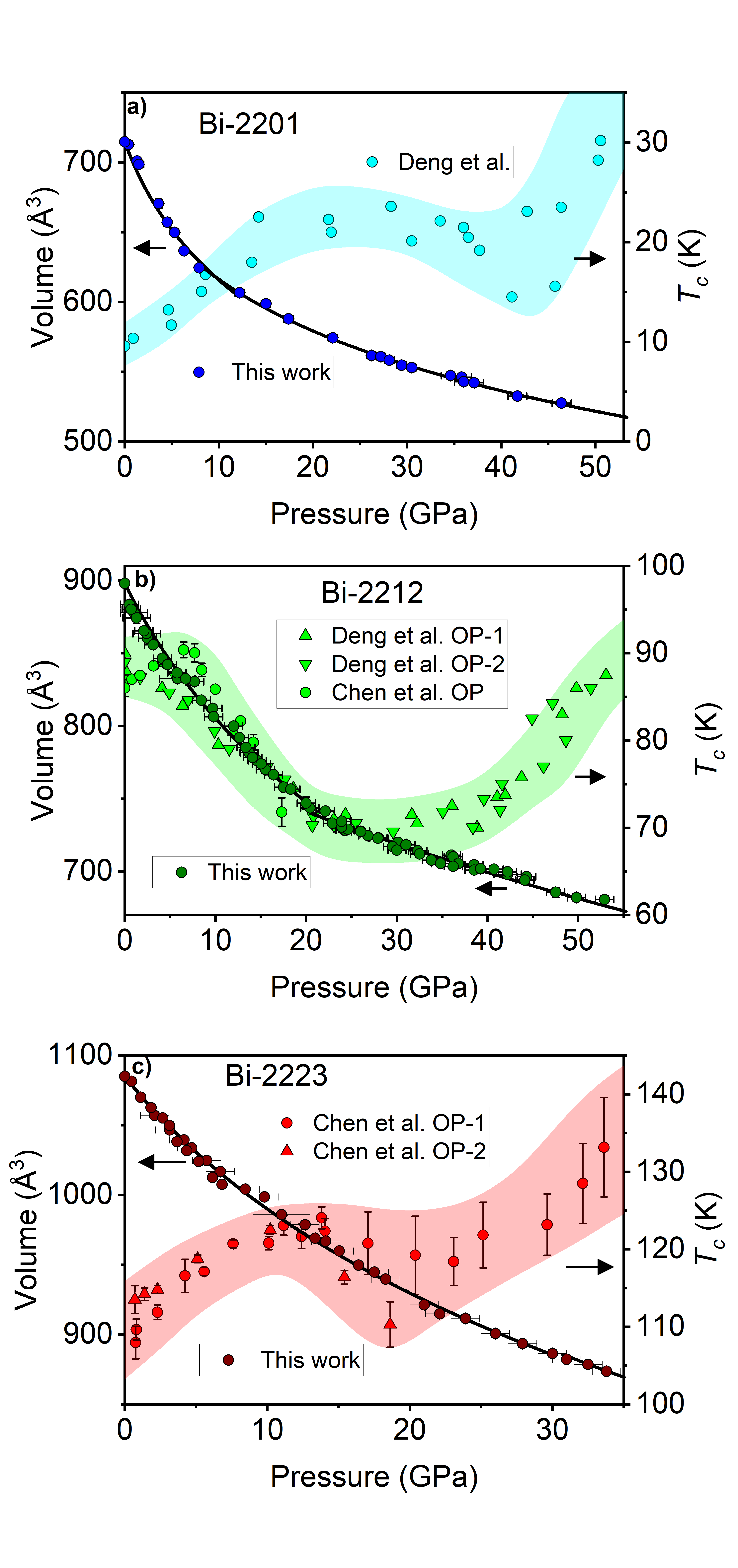}
\caption{Pressure dependence of unit-cell volume compared to previously measured $\mathit{T_c}$ for BSCCO compounds. \textbf{(a)} Bi-2201, 0.1 MPa - 8 GPa and 8 - 48 GPa EOS (see Fig. \ref{2201-measured} caption), and $\mathit{T_c}$ reported by Deng et al. \cite{deng_higher_2019}. \textbf{(b)} Bi-2212 0.1 MPa - 23 GPa and 23 - 108 GPa EOS (see Fig. \ref{2212-measured} caption), and $\mathit{T_c}$ reported by Chen et al. \cite{chen_high-pressure_2004} and Deng et al. \cite{deng_higher_2019} of Bi-2212. \textbf{(c)} Bi-2223, 0.1 MPa - 30 GPa and 30-155 GPa EOS (see Fig. \ref{2223-measured} caption), and $\mathit{T_c}$ as reported by Chen et al. \cite{chen_enhancement_2010}.}
\label{vols-tc}
\end{centering}
\end{figure}

The change in compression mechanism at 20 GPa is evident for Bi-2212 irrespective of the pressure transmitting medium used. This change is associated with a pressure-induced stiffening along the \textit{c} axis similar to that reported for other perovskite structures, such as $CaZrO_3$ \cite{yang_novel_2014} or the stiffening in \textit{b} in $CaIrO_3$ \cite{niwa_elasticity_2011}. This behavior correlates with a decrease in the magnitude of $\frac{d\mathit{T_c}}{d\mathit{P}}$ in Bi-2212 reported by Deng et al. \cite{deng_higher_2019} [Fig. \ref{vols-tc}(b)] and is consistent with the pressure-induced charge transfer (PICT) models developed to describe the effect of pressure on $\mathit{T_c}$ in the cuprates \cite{jorgensen_pressure-induced_1990}. Within these PICT models, the proximity of the rock-salt like charge reservoir layers - in particular the out of plane apical oxygen site - to the superconducting $CuO_2$ planes modify the Fermi surface in a manner similar to oxygen doping, resulting in the similar $\mathit{T_c}$-\textit{P} and $\mathit{T_c}$-doping relations. \cite{weber_apical_2010,weber_unifying_2021}.

The great impact the compression of the \textit{c} axis has on $\mathit{T_c}$ also seems to corroborate many of the recent STM experiments of O'Mahony et al. \cite{omahony_electron_2022} in which the distance of the apical oxygen site to the $CuO_2$ layers (determined by \textit{c}) greatly impacts the superconducting pairing mechanism. The fact that no structural change is apparent at pressures corresponding to the second $\mathit{T_c}$ increase in any of the materials, however,  suggests that another mechanism unrelated to the structure along \textit{c} is responsible for the phenomena. Alternate hypotheses such as high pressures suppressing a competing order \cite{chen_enhancement_2010} or inducing a Lifshitz transition \cite{deng_higher_2019} have been proposed, and more detailed high pressure studies are needed to investigate these possibilities.


Similar to the other Bi-based cuprates, fitting all Bi-2223 \textit{P}-\textit{V} data to a single Vinet EOS (0-155 GPa) results in poor fits. Splitting the fit into two pressure regions based on the kink in the pressure dependence of the \textit{c}/\textit{a} ratio at 30 GPa results in a much better fit with greatly reduced residuals. The low pressure fit produces a $\mathit{K_0} =$ 84(2) GPa and a $\mathit{K_{0}^{'}} =$ 5.6(3). The low pressure values of $\mathit{K_0}$ from the fit are significantly larger than what is measured using ultrasonic probes, which measure $\mathit{K_0}$ values between 15-40 GPa \cite{fanggao_ultrasonic_1991,dominec_elastic_1992,seetawan_synthesis_2008}. 
As with Bi-2212, $\mathit{K_{0}^{'}}$ is reported to be significantly larger when measured using ultrasonic probes, with Fanggao et al. \cite{fanggao_ultrasonic_1991-1} reporting values between 39-59 whereas our XRD measurements indicate a value for $\mathit{K_{0}^{'}}$ closer to 5.6(3). Our high pressure fit produces a $\mathit{K_0}$ value of 35.5(8) GPa with $\mathit{K_{0}^{'}} \approx$ 7.9(5).

\subsection{Stress-Strain Relations}

A stress-strain analysis was performed for each compound by linearizing the Vinet EOS to determine if any of the anomalies observed in the compression of these materials correlate to any of the other unusual pressure-induced phenomena. The linearized EOS is expressed in terms of two dimensionless parameters $\eta$, and $H(\eta)=\mathit{P}\eta ^2 / 3(1-\eta)$, and is written as

\begin{equation}
\label{eqn:lin-vinet}
\ln{(H(\eta))= \ln{(\mathit{K_0})} + \frac{3}{2}(\mathit{K'_0} - 1)(1-\eta) }.
\end{equation}

\noindent Deviations from linearity in the plot of $\ln(H(\eta))$ vs $1-\eta$ can then be used to assess pressure-induced changes in the compression properties of the material. Major changes in the trend of the stress-strain relation are present in all three materials that exhibit pressure-induced stiffening and anomalous compression mechanisms that are not well described by any of the common phenomenological EOS.

\begin{figure}[!ht]
\begin{centering}
\includegraphics[width=0.5\textwidth]{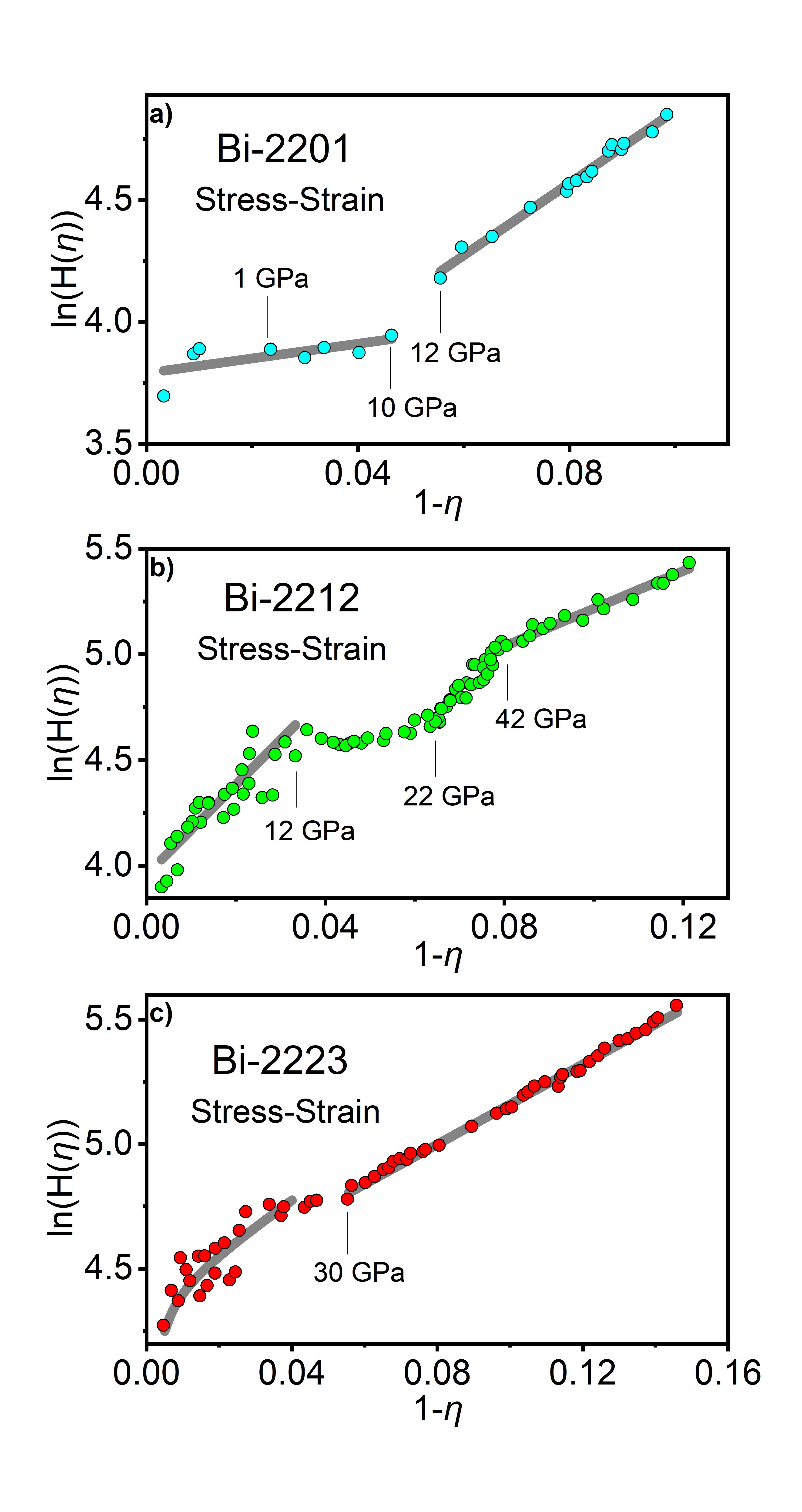}
\caption{Stress-strain relationships for all three compounds showing changes in compression mechanisms. \textbf{(a)} Bi-2201, showing a change in slope at points corresponding to pressures above 10 GPa indicating a change in compression mechanism. \textbf{(b)} Bi-2212, showing a strong deviation from linearity beginning at $12.1(1.0) GPa$, and resuming linearity at $42.2(1.0) GPa$. \textbf{(c)} Bi-2223, showing a possible kink at pressures corresponding to $23.9(1.0) GPa$ indicating a change in compression mechanism similar to Bi-2201 and Bi-2212, but much less drastic. Lines inserted as guides to the eye.} 
\label{stress-strain}
\end{centering}
\end{figure}

In Bi-2201 the stress-strain relationship is linear up to $(1-\eta) \approx 0.5$, corresponding to 10 GPa, where a discontinuity is apparent. This feature corresponds to the the pressure of the the `knee' feature in the pressure dependence of the \textit{c}/\textit{a} ratio, further reinforcing our use of multiple pressure cutoffs for fitting an EOS above and below 10 GPa. Above the discontinuity, the value of $ln(H(\eta))$ increases linearly along with the slope of the stress-strain relation [Fig. \ref{stress-strain}(a)]. 

Using a similar procedure to linearize our \textit{P}-\textit{V} data for Bi-2212 [Fig. \ref{stress-strain}(b)] results in a deviation from linearity at $(1-\eta) \approx 0.4$. Between $(1-\eta) \approx 0.4-0.8$ (12 - 42 GPa) the relation is distinctly nonlinear, indicating a major change in compression mechanism. Linearity resumes, albeit with a smaller slope, above $(1-\eta) \approx 0.8$. The midpoint of this region of non linearity corresponds to 23 GPa, the pressure of the `knee' feature apparent in the pressure response of the \textit{c}/\textit{a} ratio. The large nonlinear region indicates a degree of anomalous compression that is not accounted for in a single phenomenological EOS, with $\mathit{K_0}$ changing rapidly with pressure. Similar to Bi-2201, this result demonstrates that the compression of Bi-2212 is not well described by a single Vinet EOS over all pressure ranges.

Finally, the linearized compression data for Bi-2223 is presented in Fig. \ref{stress-strain}(c). This relation reveals a kink at $1-\eta = 0.05$ corresponding to a pressure of 30 GPa. Similar to the other two materials this kink is in agreement with the kink in the pressure dependence of \textit{c}/\textit{a}. This indicates a change in compression mechanism similar to that observed in Bi-2201 and Bi-2212, but much less pronounced. All three stress-strain relations reveal distinct high and low pressure compression regions in the three compounds.

\subsection{Anomalous Compression}

\begin{figure}[!ht]
\begin{centering}
\includegraphics[width=\textwidth]{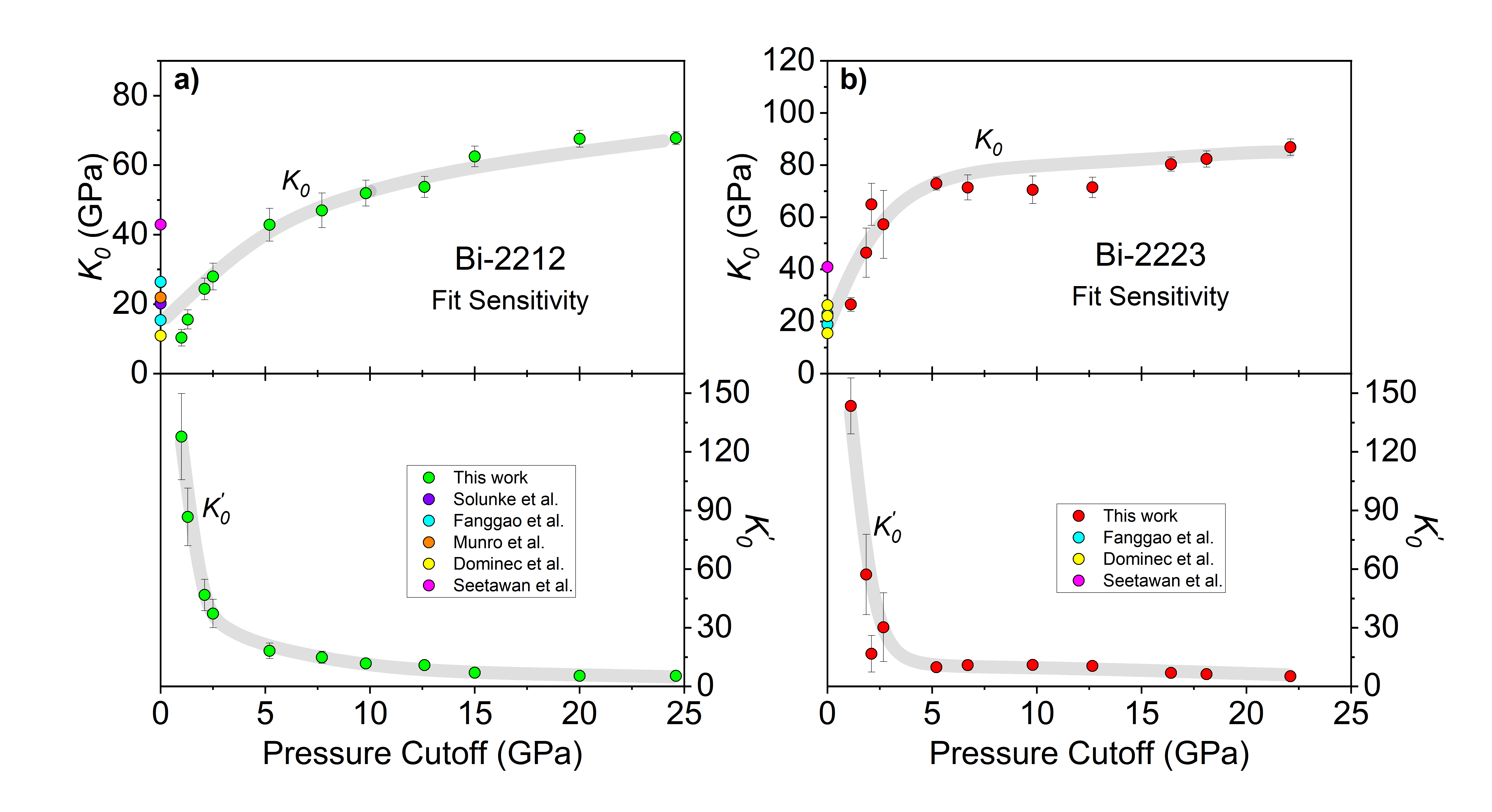}
\caption{\textbf{(a)} Top: Fit $\mathit{K_0}$ vs upper pressure cutoff used for a Vinet EOS fit to the Bi-2212 \textit{P}-\textit{V} data at pressures below the stiffening in \textit{c} presented in this work along with previously reported ambient pressure $\mathit{K_0}$ values. Bottom: Best fit values of $\mathit{K_{0}^{'}}$ for Bi-2212 for the same pressure cutoffs. \textbf{(b)} Top: Fit $\mathit{K_0}$ vs upper pressure cutoff used for a Vinet EOS fit to the Bi-2223 \textit{P}-\textit{V} data below 30 GPa along with previously reported ambient pressure $\mathit{K_0}$ values. Bottom: Best fit values of $\mathit{K_{0}^{'}}$ for Bi-2223 for the same pressure cutoffs. $\mathit{K_0}$ values from Fanggao et al. \cite{fanggao_ultrasonic_1991}, Solunke et al. \cite{solunke_effect_2007}, Munro et al. \cite{munro_elastic_2002}, and Dominec et al. \cite{dominec_elastic_1992}. Lines inserted as guides to the eye.}

\label{2212-K0}
\end{centering}
\end{figure}

Even when fitting an EOS to the \textit{P}-\textit{V} data below the obvious discontinuities in the stress-strain relations, the $\mathit{K_{0}}$ values that emerge from the fits are significantly larger than those obtained ultrasonicly. If these fitting parameters were representative of the true $\mathit{K_0}$ and $\mathit{K_{0}^{'}}$ one would expect similar values to be measured irrespective of the technique used. The analysis reveal an anomalous compression mechanism even at low pressures. This anomalous compression behavior is best exemplified in our \textit{P}-\textit{V} data for Bi-2212 and Bi-2223 below 25 GPa (Fig. \ref{2212-K0}). For both materials, fitting all data below these cutoffs results in $\mathit{K_{0}}$ values significantly larger than those $\mathit{K_{0}}$ measured in ultrasonic experiments. Reducing the upper pressure limit used in the fit results in the fitted $\mathit{K_{0}}$ value decreasing and eventually converging with the measured ultrasonic values (Fig. \ref{2212-K0}). This indicates that removing the influence of the high pressure \textit{P}-\textit{V} data results in the EOS fit producing more physically relevant parameters. The cutoff pressure dependence of $\mathit{K_{0}^{'}}$ (Fig. \ref{2212-K0}) demonstrates the significant impact of pressure on $\mathit{K_{0}^{'}}$. At low pressures the parameters that emerge from the fittings are far too large and pressure dependent, causing any model that assumes a pressure independent $\mathit{K_{0}^{'}}$ to break down upon compression. In light of the anomalous compression observed in all BSCCO compounds, any values of $\mathit{K_{0}}$ or $\mathit{K_{0}^{'}}$ obtained by a fit to an EOS should be thought of as non-physical fitting parameters that do not capture the physics present in the elastic properties of BSCCO under pressure.


\section{Conclusions}

We report structural and equation of state measurements for $Bi_2Sr_2Ca_{n-1}Cu_nO_{2n+4+\delta}$ (n=1-3) under quasi-hydrostatic compression well beyond the pressure range of previous work. These results clarify and resolve discrepancies reported in the literature, and provide information on the structural influence on the electronic structure and critical temperature of these materials. This work indicates all three BSCCO compounds to be highly susceptible to deviatoric stress, and that a structural modification does not coincide with the rise in $\mathit{T_c}$ present in all three compounds under quasi-hydrostatic compression. Additionally, we conclude that the discrepancies in the reported $\mathit{K_0}$ and $\mathit{K_{0}^{'}}$ for all three materials arise from unusual compression mechanisms beginning at very low pressures (<10 GPa) that are not well described by conventional equations of state. We propose that the anomalous high pressure behavior present in all BSCCO compounds is a manifestation of the changes in electronic properties that also give rise to the remarkable non-monotonic dependence of $\mathit{T_c}$ with pressure. These results cast high pressure studies of BSCCO compounds, and of cuprate high temperature superconductors in general, in a new light. It remains to be seen if such anomalous compression is present in other cuprate high temperature superconductors. 


\section*{Acknowledgments}

We are grateful to S.A. Gramsch and F. Restrepo for many helpful discussions, C. Li, H. Farraj, K. Kumar, and J. Cabana for assisting with the ambient pressure XRD, S. Tkachev for assisting in sample gas loading, and K. Lynch, D. Kuntzelman, R. Dojutrek, R. Frueh, J. Dublin, and M. Litz of the UIC machine shop for fabricating the high pressure cells. This work was supported by the U.S. NSF (DMR-2104881) and DOE-NNSA (DE-NA0003975, CDAC), US Air Force Office of Scientific Research Grants FA9550-15-1-236 and FA9550-20-1-0068, the T.L.L. Temple Foundation, the John J. and Rebecca Moores Endowment, and the State of Texas through the Texas Center for Superconductivity at the University of Houston (TCSUH).

\pagebreak

\section*{Figures}

\pagebreak

\pagebreak

\pagebreak

\pagebreak

\pagebreak

\pagebreak

\pagebreak

\pagebreak

\clearpage
\begingroup
\setstretch{1}
\raggedright
\printbibliography
\endgroup

\pagebreak
\end{refsection}

\begin{refsection}
\section*{Supplemental Materials}

\renewcommand{\tablename}{Table S\!\!} 
\begin{table}[!ht]
\caption{Ambient pressure lattice parameters for optimally or near-optimally doped Bi-2201 from this work (*) and reported Pryanichikov et al. \cite{pryanichnikov_nonmonotonic_2012}, Khasanova et al. \cite{khasanova_bi-2201_1995}, and Mironov et al. \cite{mironov_new_2016}. References \cite{khasanova_bi-2201_1995} and \cite{mironov_new_2016} fit the unit cell to a 4D symmetry in order to take into account the lattice supermodulation (not examined in the present work). Experimental error included when provided by source. Our \textit{a} values are multiplied by $\sqrt{2}$ in order to transform the parameters into the representation used in previous reports.}
\centering
\begin{tabular}{|c|c|c|c|c|}
\hline
\textbf{Space group} & \textbf{\textit{a} (\AA)} & \textbf{\textit{b} (\AA)} & \textbf{\textit{c} (\AA)} & \textbf{Volume (\AA$^3$)} \\
\hline
\hline
\textit{I}4/\textit{mmm}* & 5.39(1) & - & 24.6(1) & 714(3) \\
\hline
\textit{Cccm} \cite{pryanichnikov_nonmonotonic_2012} & 5.38 & 5.38 & 24.5 & 709.1 \\
\hline
\textit{P}:\textit{A}2/\textit{A}:-11 \cite{khasanova_bi-2201_1995} & 5.390(1) & 5.390(1) & 24.534(2) & 712.9(5) \\
\hline
\textit{A}12/\textit{a}1 \cite{mironov_new_2016} & 5.386(1) & 5.387(1) & 24.579(3) & 713.3(1)\\
\hline
\end{tabular}
\label{table:2201-ambient}
\end{table}

\begin{table}[!ht]
\caption{Ambient pressure lattice parameters for optimally or near-optimally doped Bi-2212 from this work (*) and reported. by Hervieu et al. \cite{hervieu_electron_1988}, Petricek et al. \cite{petricek_x-ray_1990}, Gao et al. \cite{gao_combined_1993}, Olsen et al. \cite{olsen_high_1991}, Gavarri et al. \cite{gavarri_anisotropic_1990}, Babaei et al. \cite{babaei_pour_determination_2005}, and Tarascon et al. \cite{tarascon_crystal_1988} Many of the experiments fit the unit cell to a tetragonal \textit{I}4/\textit{mmm} space group, as the orthorhombic distortion is resolution limited within a high pressure cell. All measurements were performed using XRD unless marked with a $\dagger$, indicating the data is from neutron diffraction. Experimental error included when provided by source. Our \textit{a} values are multiplied by $\sqrt{2}$ in order to transform the parameters into the representation used in previous reports.}
\centering
\begin{tabular}{|c|c|c|c|c|}
\hline
\textbf{Space group} & \textbf{\textit{a} (\AA)} & \textbf{\textit{b} (\AA)} & \textbf{\textit{c} (\AA)} & \textbf{Volume (\AA$^3$)} \\
\hline
\hline
\textit{I}4/\textit{mmm}* & 5.404(1) & - & 30.750(5) & 897.9(3) \\
\hline
\textit{Amaa} \cite{hervieu_electron_1988} & 5.405 & 5.401 & 30.715 & 896.8 \\
\hline
\textit{Amaa} \cite{petricek_x-ray_1990} & 5.408(1) & 5.413(1) & 30.871(5) & 903.7(5) \\
\hline
\textit{Amaa} \cite{gao_combined_1993} $^\dagger$ & 5.415(1) & 5.414(1) & 30.861(5) & 904.8(1) \\ 
\hline
\textit{I}4/\textit{mmm} \cite{olsen_high_1991} & 5.415(2) & - & 30.80(5) & 903.2(7) \\
\hline
\textit{Amaa} \cite{gavarri_anisotropic_1990} & 5.404(3) & 5.414(2) & 30.845(2) & 902.7(6) \\
\hline
\textit{Amaa} \cite{gavarri_anisotropic_1990} $^\dagger$ & 5.402(3) & 5.414(7) & 30.816(2) & 901.3(8) \\
\hline
\textit{Amaa} \cite{babaei_pour_determination_2005} $^\dagger$ & 5.399 & 5.405 & 30.887 & 899.3 \\
\hline
\textit{I}4/\textit{mmm} \cite{tarascon_crystal_1988} & 5.398 & - & 30.5 & 888.7 \\
\hline
\end{tabular}
\label{table:2212-ambient}
\end{table}

\begin{table}[!ht]
\caption{Ambient pressure lattice parameters for optimally or near-optimally doped Bi-2223 from this work (*) and reported by Shamray et al. \cite{shamray_crystal_2009}, Maljuk et al. \cite{maljuk_floating_2016}, and Giannini et al. \cite{giannini_growth_2005}. $\dagger\dagger$ indicates the sample was fit to a pseudo-tetragonal unit cell, that is, a tetragonal unit cell with a slight distortion in the \textit{b} parameter. Our \textit{a} values are multiplied by $\sqrt{2}$ in order to transform the parameters into the representation used in previous reports.}
\centering
\begin{tabular}{|c|c|c|c|c|}
\hline
\textbf{Space group} & \textbf{\textit{a} (\AA)} & \textbf{\textit{b} (\AA)} & \textbf{\textit{c} (\AA)} & \textbf{Volume (\AA$^3$)}  \\
\hline
\hline
\textit{I}4/\textit{mmm}* & 5.396(5) & - & 37.070(5) & 1085.0(5) \\
\hline
\textit{A}2\textit{aa} \cite{shamray_crystal_2009} & 5.411(3) & 5.409(1) & 37.082(1) & 1085.3(8) \\
\hline
\textit{I}4/\textit{mmm} \cite{maljuk_floating_2016}$^{\dagger \dagger}$ & 5.395(1) & 5.413(1) & 37.04(1) & 1081.7(7) \\
\hline
\textit{I}4/\textit{mmm} \cite{giannini_growth_2005}$^{\dagger \dagger}$ & 5.421(1) & 5.413(1) & 37.009(7) & 1086.0(4)\\
\hline
\end{tabular}
\label{table:2223-ambient}
\end{table}

\begin{table}[!ht]
\caption{$\mathit{K_0}$ and $\mathit{K_{0}^{'}}$ values from this work and reported in the literature for Bi-2201. Uncertainty included when provided. Multiple values from the same source indicate different sample runs. Due to the large probe dependent difference in $\mathit{K_0}$ values and the poor EOS fits over large pressure ranges, we believe Bi-2201 to undergo an anomalous compression mechanism that is not well described by a Vinet EOS. The $\mathit{K_0}$ and $\mathit{K_{0}^{'}}$ values determined from our data should therefore be thought of as non-physical fitting parameters and not representative of the true physical properties. Data from Seetawan et al. \cite{seetawan_synthesis_2008} and Dominec et al. \cite{dominec_elastic_1992}}
\centering
\begin{tabular}{|c|c|c|c|}
\hline
$\mathit{K_0}$ (GPa) & $\mathit{K_{0}^{'}}$ & Technique & Source \\
\hline
\hline
38.1(5.1) & 7.8(2.0) & XRD (0.1 MPa - 12 GPa) & This work \\
\hline
45.8(3.4) & 9.6(4) & XRD (8-48 GPa) & This work\\
\hline
\hline
38.6 & - & Ultrasonic & Seetawan et al. \cite{seetawan_synthesis_2008} \\
\hline
15.3 & - & Ultrasonic & Dominec et al. \cite{dominec_elastic_1992}
\\
\hline
18.9 & - & Ultrasonic & Dominec et al. \cite{dominec_elastic_1992} \\
\hline
\end{tabular}
\label{table:2201-elastic}
\end{table}

\begin{table}[!ht]
\caption{$\mathit{K_0}$ and $\mathit{K_{0}^{'}}$ values from this work and reported in the literature for Bi-2212. Uncertainty included when provided. Multiple values from the same source indicate different samples. A change in compression mechanism at 23 GPa - present in all compression experiments - means any functional fit to the \textit{P}-\textit{V} data will require two EOS. We attribute the discrepancy in the low pressure XRD and ultrasonic data to the fact that Bi-2212 compresses anomalously i.e. the pressure dependence is not captured in any of the conventional equations of state. This anomalous compression means the $\mathit{K_0}$ and $\mathit{K_{0}^{'}}$ values from fitting data to a Vinet EOS should be thought of as non-physical fitting parameters and are not representative of the true material properties. Data from Olsen et al. \cite{olsen_high_1991}, Tajima et al. \cite{tajima_pressure-effect_1989}, Zhang et al. \cite{zhang_pressure-induced_2013}, Yoneda et al. \cite{yoneda_pressure-effect_1990}, Solunke et al. \cite{solunke_effect_2007},Fanggao et al. \cite{fanggao_ultrasonic_1991-1}, Munro et al. \cite{munro_elastic_2002}, Dominec et al. \cite{dominec_elastic_1992}, and Seetawan et al. \cite{seetawan_synthesis_2008}}
\centering
\begin{tabular}{|c|c|c|c|}
\hline
$\mathit{K_0}$ (GPa) & $\mathit{K_{0}^{'}}$ & Technique & Source \\
\hline
\hline
70.5(1.8) & 4.9(3) & XRD (0.1 MPa - 23 GPa) & This work \\
\hline
262.8(22.4) & 2.9(4) & XRD (23-108 GPa) & This work \\
\hline
62(5) & 6.0(3) & XRD & Olsen et al. \cite{olsen_high_1991} \\
\hline
73 & - & XRD & Tajima et al. \cite{tajima_pressure-effect_1989} \\
\hline
127(11) & 4(fixed) & XRD & Zhang et al. \cite{zhang_pressure-induced_2013} \\
\hline
61 & - & XRD (Pb doped) & Yoneda et al. \cite{yoneda_pressure-effect_1990} \\
\hline
\hline
26.34 & - & Ultrasonic & Solunke et al. \cite{solunke_effect_2007} \\
\hline
20.18 & - & Ultrasonic & Solunke et al. \cite{solunke_effect_2007} \\
\hline
26.4 & 40 & Ultrasonic (Pb doped) & Fanggao et al. \cite{fanggao_ultrasonic_1991-1} \\
\hline
15.4 & - & Ultrasonic & Fanggao et al. \cite{fanggao_ultrasonic_1991-1} \\
\hline
21.9 & - & Ultrasonic & Munro et al. \cite{munro_elastic_2002} \\
\hline
10.9 & - & Ultrasonic & Dominec et al. \cite{dominec_elastic_1992} \\
\hline
42.89 & - & Ultrasonic & Seetawan et al. \cite{seetawan_synthesis_2008} \\
\hline

\end{tabular}
\label{table:2212-elastic}
\end{table}

\begin{table}[!ht]
\caption{$\mathit{K_0}$ and $\mathit{K_{0}^{'}}$ values from this work and reported in the literature for Bi-2223. Uncertainty included when provided. The probe dependent discrepancy in $\mathit{K_0}$ and $\mathit{K_{0}^{'}}$ values is attributed to Bi-2223 undergoing anomalous compression that is not well described by a Vinet EOS. This means the $\mathit{K_0}$ and $\mathit{K_{0}^{'}}$ values obtained through compression experiments should be thought of as fitting parameters and not physically relevant quantities. The change in compression mechanism revealed through the stress-strain relation demonstrates the need to fit the data to a high and low pressure EOS to accurately. Multiple values from the same source indicate different sample runs. Data from Yoneda et al. \cite{yoneda_pressure-effect_1990}, Fanggao et al. \cite{fanggao_ultrasonic_1991-1}, Dominec et al. \cite{dominec_elastic_1992}, and Seetawan et al. \cite{seetawan_synthesis_2008}.}
\centering
\begin{tabular}{|c|c|c|c|}
\hline
$\mathit{K_0}$ (GPa) & $\mathit{K_{0}^{'}}$ & Technique & Source \\
\hline
\hline
100.9(2.3) & 4.1(3) & XRD (0.1 MPa - 30 GPa) & This work \\
\hline
34.3(1.6) & 8.1(5) & XRD (30-155 GPa) & This work \\
\hline
73 & - & XRD & Yoneda et al. \cite{yoneda_pressure-effect_1990}\\
\hline
\hline
18.9 & 59 & Ultrasonic (Pb doped) & Fanggao et al. \cite{fanggao_ultrasonic_1991-1} \\
\hline
22.9 & 39.3 & Ultrasonic (Pb doped) & Fanggao et al. \cite{fanggao_ultrasonic_1991-1} \\
\hline
26.5 & - & Ultrasonic & Dominec et al. \cite{dominec_elastic_1992} \\
\hline
22.0 & - & Ultrasonic & Dominec et al. \cite{dominec_elastic_1992} \\
\hline
15.5 & - & Ultrasonic & Dominec et al. \cite{dominec_elastic_1992} \\
\hline
22.1 & - & Ultrasonic & Dominec et al. \cite{dominec_elastic_1992} \\
\hline
40.9 & - & Ultrasonic & Seetawan et al. \cite{seetawan_synthesis_2008} \\
\hline
\end{tabular}
\label{table:2223-elastic}
\end{table}

\newrefcontext[labelprefix=S]
\clearpage
\begingroup
\setstretch{1}
\raggedright
\printbibliography
\endgroup
\end{refsection}

\end{document}